\documentclass[smallextended]{svjour3}  

\usepackage{mathrsfs}  
\usepackage[usenames,dvipsnames]{color}
\usepackage{soul}
\usepackage{cancel}
\usepackage[
verbose,
colorlinks=true,
naturalnames=true,
linkcolor=blue,
]{hyperref}
\usepackage[normalem]{ulem}
\usepackage{graphicx}
\usepackage{subfigure}
\usepackage{amsmath,amssymb,amsfonts}
\usepackage{verbatim}
\usepackage{dcolumn}
\usepackage{bm}
\usepackage{epsf}
\usepackage{cite}

\hypersetup{colorlinks,linkcolor={blue},citecolor={blue},urlcolor={blue}} 

\newcommand{\bra}[1]{\left\langle #1\right|}
\newcommand{\ket}[1]{\left|#1\right\rangle}

\newcommand{\tr}[1]{\mathrm{tr}\left\{#1\right\}}

\newcommand{\la}{\left\langle}
\newcommand{\ra}{\right\rangle}
\newcommand{\pd}{\partial}
\newcommand{\de}[1]{\delta\left(#1\right)}

\newcommand{\e}[1]{\exp{\left(#1\right)}}
\newcommand{\lo}[1]{\ln{\left(#1\right)}}

\newcommand{\com}[2]{\left[#1,\,#2\right]}

\newcommand{\bla}{bla\\bla\\bla\\bla\\bla}

\newcommand{\mc}[1]{\mathcal{#1}}

\newcommand{\mrm}[1]{\mathrm{#1}}

\renewcommand{\vec}[1]{\boldsymbol{#1}} 

\DeclareMathOperator*{\sumint}{%
	\mathchoice%
	{\ooalign{$\displaystyle\sum$\cr\hidewidth$\displaystyle\int$\hidewidth\cr}}
	{\ooalign{\raisebox{.14\height}{\scalebox{.7}{$\textstyle\sum$}}\cr\hidewidth$\textstyle\int$\hidewidth\cr}}
	{\ooalign{\raisebox{.2\height}{\scalebox{.6}{$\scriptstyle\sum$}}\cr$\scriptstyle\int$\cr}}
	{\ooalign{\raisebox{.2\height}{\scalebox{.6}{$\scriptstyle\sum$}}\cr$\scriptstyle\int$\cr}}
}

\begin{document}
	\title{Jarzynski equality for conditional stochastic work}

	\author{Akira Sone \and Sebastian Deffner}
	
	\institute{Akira Sone \at
		Theoretical Division, Los Alamos National Laboratory, Los Alamos, NM 87545, USA \\
		Center for Nonlinear Studies, Los Alamos National Laboratory, Los Alamos, NM 87545, USA\\
		\email{akirasone628@gmail.com}  
		\and
		Sebastian Deffner \at
		Department of Physics, University of Maryland, Baltimore County, Baltimore, MD 21250, USA\\
		Instituto de F\'isica `Gleb Wataghin', Universidade Estadual de Campinas, 13083-859, Campinas, S\~{a}o Paulo, Brazil\\
		\email{deffner@umbc.edu} 
	}
	
	\date{Received: date / Accepted: date}
	
	\maketitle
	\begin{abstract}
		It has been established that the \emph{inclusive work} for classical, Hamiltonian dynamics is equivalent to the two-time energy measurement paradigm in isolated quantum systems. However, a plethora of other notions of quantum work has emerged, and thus the natural question arises whether any other quantum notion can provide motivation for purely classical considerations. In the present analysis, we propose the \emph{conditional stochastic work} for classical, Hamiltonian dynamics, which is inspired by the one-time measurement approach. This novel notion is built upon the change of expectation value of the energy conditioned on the initial energy surface. As main results we obtain a generalized Jarzynski equality and a sharper maximum work theorem, which account for how non-adiabatic the process is. Our findings are illustrated with the parametric harmonic oscillator.
	\end{abstract}

	\section{Introduction}
	\label{sec:intro}
	
	Over the last two decades stochastic thermodynamics \cite{Seifert2008,Jarzynski2015,Broeck2016} has developed  into  arguably the most active field in modern statistical physics. This development was initiated by the discovery of the fluctuation theorems \cite{Evans1993,Gallavotti1995,Evans2002,Zon2003,Sevik2008,Shargel2010}, which are statements of the second law quantifying the occurrence of negative fluctuations of the entropy production. Chief among these theorems is the Jarzynski equality \cite{Jarzynski97}, which despite its innocuous appearance might be one of the most powerful results in thermodynamics discovered to date \cite{Ortiz2011}. It is typically written as 
	\begin{equation}
		\label{eq:jar}
		\la \e{-\beta W}\ra=\e{-\beta \Delta F}\,,
	\end{equation}
	where $\beta=1/k_B T$ is the inverse temperature, $W$ is the work performed by the system of interest, and $\Delta F$ is the free energy difference. The average $\la\cdots\ra$ is taken over an ensemble of realizations of the same process induced by the variation of some external control parameter $\alpha$. 
	
	The Jarzynski equality is a universally valid theorem \cite{Jarzynski97,Jarzynski97pre,Jarzynski2004,Chernyak2005,Chernyak2006,Speck2007,Jarzynski2008} that holds in any classical system, as long as the initial state as well as an instantaneous stationary state of the dynamics is described by a thermal Gibbs distribution $p^\mrm{eq}(\Gamma,\alpha)=\e{-\beta H(\Gamma,\alpha)}/Z(\alpha)$, where $H(\Gamma, \alpha)$ is the Hamiltonian, $\Gamma$ denotes a point in phase space, and $Z(\alpha)$ is the (canonical) partition function. Its importance can hardly be overestimated once one realizes that Eq.~\eqref{eq:jar} replaces the \emph{inequality} of the conventional maximum work theorem, $\la W\ra\geq\Delta F$, by an \emph{equality}. Hence, the Jarzynski equality provides means to compute the difference of free energy, i.e., an equilibrium property, from an ensemble of non-equilibrium realizations of the same process. Interestingly, different notions of thermodynamic work $W$ have been considered in the literature  that obey their own variants of Eq.~\eqref{eq:jar}. The most prominent examples are the  \emph{exclusive}  \cite{Bochkov1981} and \emph{inclusive} work \cite{Horowitz07,Jarzynski07}, of which Jarzynski himself proposed the latter.
	
	Therefore, Eq.~\eqref{eq:jar} has also been the main cornerstone in the quest to generalize thermodynamic notions to the quantum domain \cite{DeffnerBook19,Binder19}. However, finding physically meaningful and experimentally relevant notions of quantum work is a rather involved task. To date, a zoo of different approaches has been proposed \cite{Campisi2013,Deffner2013,Allahverdyan2014,Talkner2016,Diaz20,Levy2020,Micadei2020,Beyer2020}, of which the so-called two-time energy measurement paradigm \cite{Tasaki00,Kurchan01,Talkner2007} is probably the most widely used in open and closed systems \cite{Jarzynski04,Morikuni17,Rastegin13,Jarzynski15,Zhu16,Funo18,Pan19,Kafri12,Albash13,Goold15,Rastegin14,Goold14,Marti17,Gardas2018}. Whether or not a notion of quantum work is considered ``reasonable" is typically assessed by testing whether a quantum version of the Jarzynski equality can be derived \cite{Hanggi2015}. 
	
	Given how important the Jarzynski equality and the corresponding classical notion of stochastic work have been in the development of quantum thermodynamics, it is thus only a fair question to ponder whether quantum thermodynamics can return the favor. The present analysis is motivated by the one-time measurement paradigm, which was proposed for isolated systems in Ref.~\cite{Deffner16}, and then generalized to open systems \cite{Sone20a}. This paradigm is designed to assess the informational contribution to the maximum work theorem due to the quantum back action of the projective measurements in the two-time measurement paradigm. The main insights from the one-time measurement paradigm \cite{Deffner16} are (i) a new notion of quantum work that relies only on the expected energy and (ii) a sharpened maximum work theorem that includes a term accounting for how far from equilibrium a quantum system is driven.
	
	In the following, we develop a classical equivalent of the one-time measurement paradigm. To this end, we introduce the notion of \emph{conditional stochastic work}, which is based on identifying work as the change of conditional expectation values of the energy. As main result we obtain an integral fluctuation theorem, which can be interpreted as a generalized Jarzynski equality, and which leads to a tightened maximum work theorem. The analysis is concluded with a pedagogical example, namely the parametric harmonic oscillator whose quantum version was studied in Ref.~\cite{Deffner16}.
	
	\section{Motivation: quantum work from one-time measurements}
	
	We start by briefly summarizing the  quantum paradigms, namely the two- and one-time measurement paradigms, and by establishing notions and notations. For the two-time measurement paradigm we particularly focus on the quantum-classical correspondence \cite{Jarzynski15}, which provides a natural template for the definition of the \emph{conditional stochastic work}.
	
	\subsection{Quantum-classical correspondence for two-time measurement approach}
	
	\paragraph{Quantum work in the two-time measurement paradigm}
	
	For the purposes of our present analysis we restrict ourselves to isolated system\footnote{The two-time \cite{Kafri12,Albash13,Rastegin13,Smith2018,Gardas2018} as well as the one-time \cite{Sone20a} measurement paradigm have been extended to open system dynamics. Thus, a generalization of our present discussion to open systems appears straight forward.}. To this end, consider a quantum system whose dynamics is described by the driven von Neumann equation,
	\begin{equation}
		\label{eq:vN}
		i\hbar \frac{d}{dt}\,\rho(t)=\com{H(\alpha_t)}{\rho(t)}\,,
	\end{equation}
	where $\alpha_t$ is the external control parameter that is varied from $\alpha_0$ to $\alpha_\tau$ during time $\tau$. Taking inspiration from Ref.~\cite{Jarzynski15}, we denote the initial Hamiltonian by $H(\alpha_0)\equiv H_A$, and the final Hamiltonian by $H(\alpha_\tau)\equiv H_B$.
	
	At $t=0$ the system is prepared in a Gibbs state, $\rho_A^\mrm{eq}=\e{-\beta H_A}/Z_A$, at inverse temperature $\beta$. Then a projective measurement of the energy is taken, before the system is let to evolve under Eq.~\eqref{eq:vN}, before  a second projective measurement is taken at $t=\tau$. Then, the quantum work is defined as difference of measurement outcomes, i.e., difference of initial and final energy eigenvalues
	\begin{equation}
		W_\mrm{ttm}\left(\ket{m_A}\rightarrow\ket{n_B}\right)=E_n^B-E_m^A\,.
	\end{equation}
	Its distribution is given as an average over all realizations
	\begin{equation}
		\mc{P}_\mrm{ttm}(W)=\la \de{W-W_\mrm{ttm}(\ket{m_A}\rightarrow\ket{n_B})}\ra\,,
	\end{equation}
	which can be written explicitly as
	\begin{equation}
		\label{eq:pTTM}
		\mc{P}_\mrm{ttm}(W)=\sumint_{n,m} \de{W-E_n^B+E_m^A}\, p\left(\ket{m_A}\rightarrow\ket{n_B}\right)\,.
	\end{equation}
	Here, we use the notation $\sumint$ to represent a sum over the discrete and continuous parts of the eigenvalues. 	In the two-time measurement paradigm the probability for single ``realizations'' reads \cite{Kafri12},
	\begin{equation}
		p\left(\ket{m_A}\rightarrow\ket{n_B}\right)=\tr{\Pi_n^B\, U_\tau\, \Pi_m^A \rho_A^\mrm{eq} \Pi_m^A\, U_\tau^\dagger}\,,
	\end{equation}
	where $\Pi_m^A$ and $\Pi_n^B$ are the projectors into the eigenbasis of $H_A$ and $H_B$, respectively. Further, $U_\tau=\mc{T}_> \e{-i/\hbar\,\int_0^\tau dt\,H(\alpha_t)}$ is the evolution operator corresponding to Eq.~\eqref{eq:vN}. For non-degenerate spectra, $p\left(\ket{m_A}\rightarrow\ket{n_B}\right)$ can be written as
	\begin{equation}
		\label{eq:ptqm}
		p\left(\ket{m_A}\rightarrow\ket{n_B}\right)=p_{n,m}^\tau p_m^A=\left|\bra{n_B}U_\tau\ket{m_A}\right|^2\,p_m^A\,,
	\end{equation}
	where $p_m^A=\e{-\beta E_m^A}/Z_A$ are the thermal occupation probabilities, and $p_{n,m}^{\tau}=\left|\bra{n_B}U_\tau\ket{m_A}\right|^2$ is the quantum transition probability.
	
	For this notion of quantum work, it is a simple exercise to derive the Jarzynski equality \eqref{eq:jar}, which essentially follows from the properties of the unitary dynamics \cite{Talkner07,DeffnerBook19}. More generally the two-time measurement paradigm provides a \emph{proper} notion of quantum work also in open systems if the dynamics is unital \cite{Kafri12,Rastegin13,DeffnerBook19,Smith2018}.
	
	\paragraph{Classical, inclusive work}
	
	Interestingly, the two-time measurement paradigm can be ``translated'' directly to the classical domain. The dynamics of the corresponding classical scenario is described by the Liouville equation,
	\begin{equation}
		\label{eq:liou}
		\frac{d}{dt} p(\Gamma, t)=\{H(\Gamma; \alpha_t),p(\Gamma,t)\}\,,
	\end{equation}
	where $\{.,.\}$ is the Poisson bracket, and $\Gamma \equiv(\vec{q}_1,\vec{p}_1,\cdots,\vec{q}_n,\vec{p}_n) $ is a point in phase space. The classical (inclusive) work is then defined by  the difference of the Hamiltonians at $t=0$ and $t=\tau$ \cite{Jarzynski97},
	\begin{equation}
		W_\mrm{class}(E_A\rightarrow E_B)=E_B-E_A=H(\Gamma_\tau(\Gamma_0); \alpha_\tau)-H(\Gamma_0; \alpha_0)\,,
	\end{equation}
	where $\Gamma_0$ denotes the locus of phase space points on the surface of energy $E_A$ . Accordingly, $\Gamma_\tau(\Gamma_0)$ denotes the final locus of a trajectory that started on the initial energy surface, $H(\Gamma_0; \alpha_0)=E_A$, and ended at $H(\Gamma_\tau(\Gamma_0); \alpha_\tau)=E_B$. 
	
	The classical work distribution is then given by
	\begin{equation}
		\mc{P}_\mrm{class}(W)=\la \de{W-W_\mrm{class}(E_A\rightarrow E_B}\ra\,,
	\end{equation}
	which can be written more explicitly as
	\begin{equation}
		\label{eq:pclass}
		\mc{P}_\mrm{class}(W)=\int dE_A\int dE_B\,\de{W-E_B+E_A} p(E_B|E_A)\, p^\mrm{eq}(E_A)\,.
	\end{equation}
	In complete analogy to the quantum case, $p^\mrm{eq}(E_A)=\e{-\beta E_A}/Z_A $ denotes the initial, thermal distribution of energy shell $E_A$.  The conditional ``transition'' probability distribution $p(E_B|E_A)$ can be written as \cite{Jarzynski15}
	\begin{equation}
		\label{eq:ptclass}
		p(E_B|E_A)=\frac{\int d\Gamma_0\,\de{E_A-H(\Gamma_0; \alpha_0)}\,\de{E_B-H(\Gamma_\tau(\Gamma_0); \alpha_\tau)}}{\int d\Gamma_0\,\de{E_A-H(\Gamma_0; \alpha_0)}}\,,
	\end{equation}
	which is a continuum limit of the quantum transition probability \eqref{eq:ptqm}.
	
\begin{figure}
		\centering
		\includegraphics[width=.9\columnwidth]{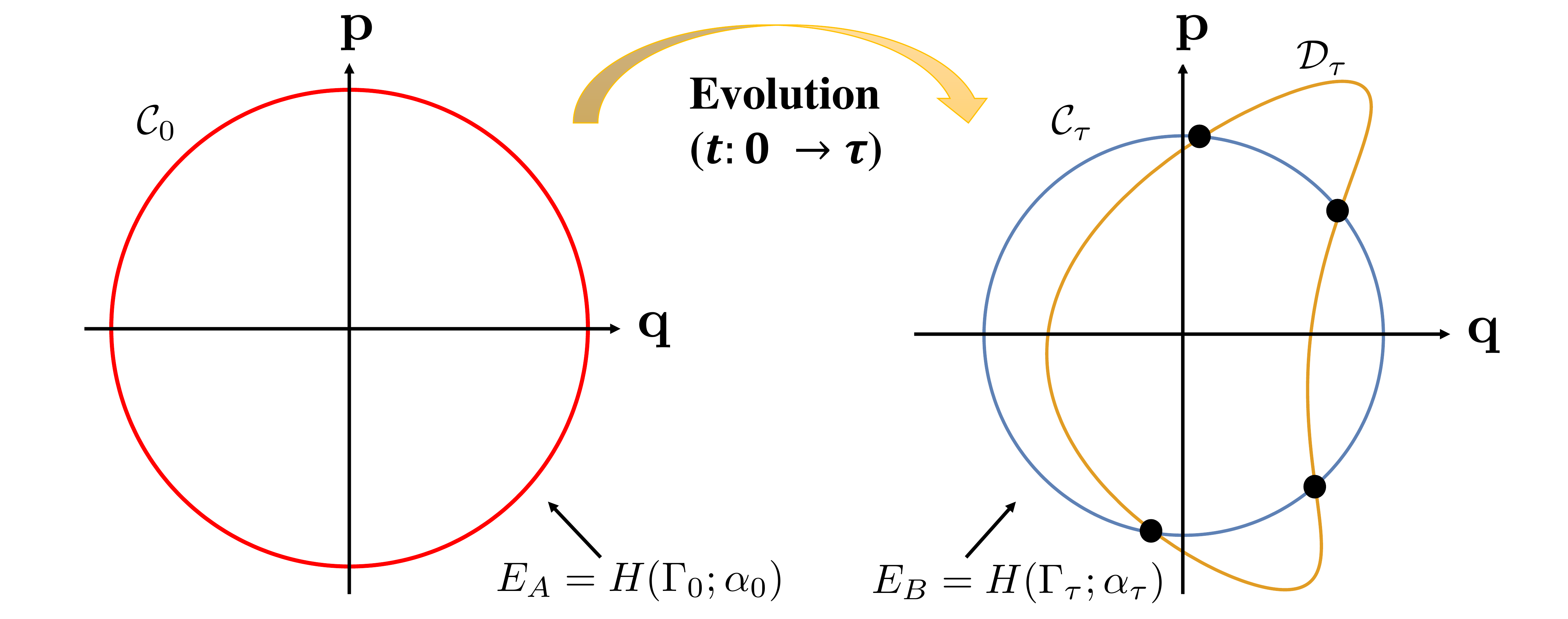}
		\caption{\label{fig1} An energy shell $\mc{C}_0$ of the initial Hamiltonian, $H(\Gamma_0; \alpha_0)$,  evolves under the Liouville equation \eqref{eq:liou} to the curve $\mc{D}_\tau$. The curve $\mc{C}_\tau$ is an energy shell of the final Hamiltonian, $H(\Gamma_\tau; \alpha_\tau)$. The transition probability distribution \eqref{eq:ptclass} counts the intersection points of $\mc{D}_\tau$ with $\mc{C}_\tau$. }
	\end{figure}

	In Ref.~\cite{Jarzynski15} it was then shown that the quantum work distribution \eqref{eq:pTTM} converges towards the classical work distribution \eqref{eq:pclass} in the semiclassical limit $\hbar\ll1$. More importantly, however, we emphasize that the classical transition probability distribution \eqref{eq:ptclass} has a similar interpretation as the quantum transition probability \eqref{eq:ptqm}. Whereas $p_{n,m}^\tau$ \eqref{eq:ptqm} quantifies the overlap of the time-evolved initial eigenstates, $U\ket{m_A}$, with the final energy eigenstates, $\ket{n_B}$, the classical expression \eqref{eq:ptclass} counts the intersection points of the time-evolved initial energy shell, $\mc{D}_\tau$, and the final energy shell, $\mc{C}_\tau$ (see Fig.~\ref{fig1}).

	\subsection{Quantum work from the one-time measurement paradigm}
	
	In the following, we will develop a classical analogue of the one-time measurement paradigm. This approach was motivated by the desire to better quantify the thermodynamic contribution of the measurement back action \cite{Deffner16}. To this end, quantum work is determined by (i) preparing the quantum system in a thermal Gibbs state, (ii) taking an initial, projective measurement of the energy at $t=0$, and then (iii) letting the system evolve under the unitary dynamics \eqref{eq:vN}. Note that the protocol is identical to the two-time measurement paradigm, except that no measurement is taken at $t=\tau$.
	
	In this case, quantum work is given by \cite{Deffner16}
	\begin{equation}
		\label{eq:Wotm}
		W_\mrm{otm}\left(\ket{m_A}\right)=\bra{m_A} U^\dagger_\tau\,H_B\,U_\tau\ket{m_A}-E_m^A\,.
	\end{equation}
	Hence, quantum work for a single realization is determined by considering how much the expectation value for a single energy eigenstate changes under the unitary evolution. It is easy to see \cite{Deffner16} that $\la W_\mrm{otm}\ra=\la W_\mrm{ttm}\ra$. Therefore, the one-time measurement paradigm is uniquely suited to assess the thermodynamics of measurements, which manifests in statements of the second law \cite{Deffner16}.
	
	The corresponding work distribution becomes
	\begin{equation}
		\mc{P}_\mrm{otm}(W)=\la \de{W-W_\mrm{otm}\left(\ket{m_A}\right)}\ra\,,
	\end{equation}
	which can be written for non-degenerate spectra as
	\begin{equation}
		\mc{P}_\mrm{otm}(W)=\sumint_{m} \de{W-\bra{m_A} U^\dagger_\tau\,H_B\,U_\tau\ket{m_A}+E_m^A}\, p_m^A\,.
	\end{equation}
	
	It is important to note that in the one-time measurement paradigm all information about the time-evolution is encoded in the work values, and that the average is only taken over the initial distribution. In contrast, in the two-time measurement paradigm the work values are taken as differences of energy eigenvalues, and the time-dependence of the problem is entirely expressed through the transition probability distribution. This will become important in the following, where we construct the classical analogue of the one-time measurement paradigm.

	\section{Fluctuation theorem for conditional stochastic work}
	\label{sec:result}
	
	\subsection{Time-evolved energy shells -- conditional stochastic work}
	
	We continue by defining the \emph{conditional stochastic work}. In complete analogy to the quantum case \eqref{eq:Wotm}, we consider the change in conditional expectation values of the energy under the Liouville dynamics \eqref{eq:liou}. Thus, we have
	\begin{equation}
		\label{eq:Wps}
		W_\mrm{c}(E_A)\equiv \int dE_B\,E_B\,p(E_B|E_A)-E_A=\varepsilon_B(E_A)-E_A\,,
	\end{equation}
	where $p(E_B|E_A)$ is again the classical transition probability distribution \eqref{eq:ptclass}, and where we introduced the notation 
	\begin{equation}
		\varepsilon_B(E_A)\equiv\int dE_B\,E_B\,p(E_B|E_A)\,.
		\label{eq:epsilon}
	\end{equation}
	In other words, we define the work as the difference between the conditional expectation value $\varepsilon_B(E_A)$ at time $t=\tau$ and the initial energy $E_A$, given that the system started on the energy surface $E_A=H(\Gamma_0; \alpha_0)$.
	
	Then, the probability distribution of  the such defined $W_\mrm{c}(E_A) $ reads
	\begin{equation}
		\mc{P}_\mrm{c}(W)=\int dE_A \, \de{W-\varepsilon_B(E_A)+E_A}\,p^\mrm{eq}(E_A)\,,
	\end{equation}
	where as before $p^\mrm{eq}(E_A)$ is the thermal distribution over the initial energy surfaces. Therefore, as above in the quantum case, the signatures of the dynamics are encoded in the work values, whereas $\mc{P}_\mrm{c}(W)$ is determined as an average over the initial distribution only. 
	
	We also immediately observe that
	\begin{equation}
		\label{eq:Wav}
		\la W\ra_\mrm{c}=\int d W\, W\, \mc{P}_\mrm{c} (W)=\int d W\, W\, \mc{P}_\mrm{class} (W)=\la W \ra_\mrm{class}=\la E_B\ra-\la E_A\ra \,.
	\end{equation}
	Hence, the average conditional stochastic work is identical to the average classical, inclusive work, which is simply given by the difference of the average final and initial energies. Thus, the conditional stochastic work is fully consistent with the first law of thermodynamics for isolated systems, and can hence be considered a \emph{good} notion of thermodynamic work \cite{DeffnerBook19}.

	\subsection{Generalized Jarzynski equality for conditional stochastic work}
	
	To derive the generalized Jarzynski equality we now compute the average of the exponentiated work. We have
	\begin{equation}
		\la \e{-\beta W}\ra_\mrm{c}=\int dW\, \e{-\beta W}\,\mc{P}_\mrm{c}(W)\,,
	\end{equation}
	and we immediately obtain
	\begin{equation}
		\label{eq:FT1}
		\la \e{-\beta W}\ra_\mrm{c}=\frac{1}{Z_A}\,\int dE_A\,\e{-\beta\, \varepsilon_B(E_A) }\,.
	\end{equation}
	The latter can be interpreted as the ratio of two partition functions, namely the canonical partition function of the initial distribution $Z_A$, and 
	\begin{equation}
		\label{eq:Zcond}
		\mc{Z}(B|A)=\int dE_A\,\e{-\beta\, \varepsilon_B(E_A)} \,.
	\end{equation}
	Here, $\mc{Z}(B|A)$ is the partition function of the conditional thermal distribution, 
	\begin{equation}
		p_{B|A}(E_A,E_B) \equiv p(E_B|E_A)\,\frac{\e{-\beta\, \varepsilon_B(E_A)}}{\mc{Z}(B|A)}\,,
		\label{eq:CondThermalDis}
	\end{equation}
	which is the thermal distribution of the conditional expectation values of the final energy corresponding to the initial energy surfaces. The distribution \eqref{eq:CondThermalDis} is the classical analogue of what has been dubbed ``best guessed state'' in the quantum context \cite{Deffner16}. Here, note that $p_{B|A}(E_A,E_B)$ is a joint probability distribution of $E_A$ and $E_B$. We use the word ``conditional" because it is associated with the conditional expectation value of the final energy.
	
	Therefore, Eq~\eqref{eq:FT1} can be further re-written as
	\begin{equation}
		\label{eq:FT2}
		\la \e{-\beta W}\ra_\mrm{c}=\e{-\beta \Delta F}\,\frac{\mc{Z}(B|A)}{Z_B}\,,
	\end{equation}
	where the free energy is as usual, $F=-1/\beta\, \lo{Z}$, and $\Delta F=F_B-F_A$. Equation~\eqref{eq:FT2} is almost in the form of the desired result, namely a generalized Jarzynski equality \eqref{eq:jar}. However, the ratio of the conditional and canonical partition function needs a clearer physical interpretation. Therefore, we now consider the Kullblack-Leiber (KL) divergence \cite{Kullback1951} of the conditional thermal distribution \eqref{eq:CondThermalDis}  with respect to the canonical state of the final configuration,
	\begin{equation}
		D\left[p_{B|A}\,||\,p^\mrm{eq}_B\right]=\int dE_B \int dE_A\, p_{B|A}(E_A,E_B)\,\lo{\frac{p_{B|A}(E_A,E_B)}{p^\mrm{eq}(E_B)}}\,.
	\end{equation}
	This KL-divergence is the difference of (minus) the Shannon entropy of the conditional distribution \eqref{eq:CondThermalDis}, and the cross entropy of conditional and canonical distributions. From Eqs.~\eqref{eq:CondThermalDis} and $p_B^{\mrm{eq}}(E_B)=\e{-\beta E_B}/Z_B$, we obtain
	\begin{equation}
		\label{eq:KL}
		D\left[p_{B|A}\,||\,p^\mrm{eq}_B\right]=-\lo{\frac{\mc{Z}(B|A)}{Z_B}}\,,
	\end{equation}
	which expresses that the ratio of the partition functions directly quantifies the distinguishability of the conditional and canonical distributions. This also means that $\mc{Z}(B|A)/Z_B$ measures ``how far from equilibrium'' the system is driven by the variation of $\alpha_t$. A proof of Eq.~\eqref{eq:KL} is outlined in Appendix \ref{sec:proofKL}.
	
	The last point becomes even clearer in the generalized Jarzynski equality. By employing the KL-divergence \eqref{eq:KL}, Eq.~\eqref{eq:FT2} can then be written as
	\begin{equation}
		\label{eq:FT3}
		\la \e{-\beta W}\ra_\mrm{c}=\e{-\beta \Delta F-D\left[p_{B|A}\,||\,p^\mrm{eq}_B\right]}\,,
	\end{equation}
	which constitutes our main result. Using the standard argument, namely the convexity of the exponential and Jensen's inequality we further have
	\begin{equation}
		\label{eq:Wmax}
		\beta \la W\ra \geq \beta\Delta F+D\left[p_{B|A}\,||\,p^\mrm{eq}_B\right]\,,
	\end{equation}
	which is a \emph{tighter} version of the maximum work theorem. Note again that $\la W \ra_\mrm{c}=\la W \ra_\mrm{class}=\la W\ra$ \eqref{eq:Wav}. Hence, Eq.~\eqref{eq:Wmax} expresses not only that the average, finite-time work is larger than the free energy difference, but also that is actually larger than the free energy difference \emph{plus} a positive term quantifying the non-adiabaticity of the process. This term arises from conditioning the single work values on the initial energy surface, and considering the change of the conditional expectation value. For truly equilibrium processes this term vanishes $D\left[p_{B|A}\,||\,p^\mrm{eq}_B\right]=0$. It is interesting to note that the present discussion can be generalized to open systems. In this case, $E_A$ and $E_B$ are the initial total energy and possible final energies of the universe. Thus, we could obtain the same formula, where the conditional thermal state and Gibbs state in the KL-divergence are associated with the total system. However, a more careful treatment of open system dynamics requires a more thorough analysis, which is outside the scope of the present work.

	\section{Illustrative example: parametric harmonic oscillator}
	\label{sec:example}
	
	We conclude the analysis with an illustrative case study, namely the one-dimensional parametric harmonic oscillator. The Hamiltonian reads,
	\begin{equation}
		\label{eq:harm}
		H(q,p;\omega_{t}) = \frac{p^2}{2m}+\frac{1}{2}\,m\omega_t^2\,q^2\,,
	\end{equation}
	where $\omega_t \geq 0$ and $m$ is the mass. For our present purposes this system is particularly well-suited since its dynamics is analytically solvable \cite{Husimi53}, and its thermodynamic properties have been analyzed for classical systems \cite{Jarzynski97pre}, quantum two-time \cite{Zon2008,Deffner08,Huber2008,Deffner2010CP,Deffner2013PRE,Gong2014,Deng17b,Myers2020} and one-time measurement paradigms \cite{Deffner16}.
	
	The situation can be conveniently analyzed with the mathematical tool kit put forward in Refs.~\cite{Deng17a,Deng17b}. To this end, we write the transition probability distribution \eqref{eq:ptclass} in terms of the phase space volume up to an energy $E$,
	\begin{equation}
		\Omega(E; \alpha)=\int d\Gamma \,\Theta\left(E-H(\Gamma; \alpha)\right)\,,
	\end{equation}
	where $\Theta(\cdot)$ is the Heaviside step function. Noting that the relation between energy and phase space volume is a one-to-one map, we can also write $E=E(\Omega; \alpha)$. Thus, the transition probability distribution \eqref{eq:ptclass} can also be expressed as \cite{Deng17a}
	\begin{equation}
		\label{eq:ptOmega}
		p(\Omega_\tau|\Omega_0)=\frac{\int d\Gamma_0\,\de{E(\Omega_0; \alpha_0)-H(\Gamma_0; \alpha_0)}\,\de{\Omega_\tau-\Omega(H(\Gamma_\tau(\Gamma_0); \alpha_\tau); \alpha_\tau)}}{\int d\Gamma_0\,\de{E(\Omega_0; \alpha_0)-H(\Gamma_0; \alpha_0)}}\,,
	\end{equation}
	where $\Omega_0$ is the initial phase space volume, and $\Omega_\tau$ is the volume evolved under the Liouville equation \eqref{eq:liou}. A proof of the equivalence of the expressions in terms of energies and phase space volume is summarized in Appendix \ref{sec:TransProb}.
	
	For the parametric harmonic oscillator, the external control parameter is identified with the angular frequency $\alpha=\omega$, and we simply have
	\begin{equation}
		\label{eq:omega}
		\Omega(E; \omega)=\frac{E}{\hbar\omega}\quad\text{and}\quad E(\Omega; \omega)= \hbar \omega\,\Omega\,.
	\end{equation}
	Moreover, the normalization of $p(\Omega_\tau|\Omega_0)$, i.e., the density of states \cite{Deng17a} becomes
	\begin{equation}
		\label{eq:dens}
		\int d\Gamma\,\de{E-H(\Gamma; \omega)}=\frac{1}{\hbar\omega}\,.
	\end{equation}
	
	To compute $p(\Omega_\tau|\Omega_0)$ \eqref{eq:ptOmega} we now need the solution of the time-dependent dynamics. This can be written as \cite{Husimi53}
	\begin{equation}
		q_t=Y_t\,q_0+X_t\,\frac{p_0}{m}\quad\text{and}\quad    p_t=m\dot{Y}_t\,q_0+\dot{X}_t\,p_0\,,
	\end{equation}
	where the dot denotes a derivative with respect to time. Here, the $X_t, Y_t$ are particular solutions of the force-free equation of motion $\ddot{X}+\omega^2(t) X=0$, satisfying $X_0=\dot{Y}_0=0$ and $\dot{X}_0=Y_0=1$~\cite{Husimi53}. They further obey for all $t$,
	\begin{equation}
		X_t\dot{Y}_t-\dot{X}_tY_t=1\,.
		\label{eq:XandY}
	\end{equation}
	
	We now define the Hamiltonian flow vector
	\begin{equation}
		\vec{v}=\left(\frac{p_0}{\sqrt{m\omega_0}},\sqrt{m\omega_0}q_0\right)^T\,,
	\end{equation} 
	with which we can write 
	\begin{equation}
		p(\Omega_{\tau}|\Omega_{0})=\int d\Gamma \de{\Omega_{\tau}-\pi\,\vec{v}^{T}\vec{M}\vec{v}}\,\de{\Omega_{0}-\pi\,\vec{v}^T\vec{v}}\,,
	\end{equation}
	and where we used Eqs.~\eqref{eq:omega} and \eqref{eq:dens}. Here, $\vec{M}$ is the $2\times2$ symmetric matrix
	\begin{equation}
		\begin{split}
			\vec{M} = 
			\begin{pmatrix}
				X_{\tau}^2\omega_{0}\omega_{\tau}+\dot{X}_{\tau}^2\frac{\omega_{0}}{\omega_{\tau}}&\frac{1}{\omega_{\tau}}\dot{X}_{\tau}\dot{Y}_{\tau}+X_{\tau}Y_{\tau}\omega_{\tau}\\
				\frac{1}{\omega_{\tau}}\dot{X}_{\tau}\dot{Y}_{\tau}+X_{\tau}Y_{\tau}\omega_{\tau}&\frac{1}{\omega_{0}\omega_{\tau}}\dot{Y}_{\tau}^2+Y_{\tau}^2\frac{\omega_{\tau}}{\omega_0}
			\end{pmatrix}\,.
		\end{split}
	\end{equation}
	The trace of $\vec{M}$ is directly related to the measure of adiabaticity of the process, $Q^*$. It is defined as \cite{Husimi53},
	\begin{equation}
		Q^{*} = \frac{1}{2\omega_{0}\omega_{\tau}}\left[\omega_{0}^2\left(\omega_{\tau}^2 X_{\tau}^2+\dot{X}_{\tau}^2\right)+\left(\omega_{\tau}^2 Y_{\tau}^2+\dot{Y}_{\tau}^2\right)\right]\,,
		\label{eq:Q}
	\end{equation}
	and we have 
	\begin{equation}
		Q^{*}=\frac{1}{2}\,\tr{\vec{M}}\,.
	\end{equation}
	Note that $Q^{*}=1$ for adiabatic process, and $Q^{*}>1$ for non-adiabatic process~\cite{Husimi53}.
	
	Now, let $\mu_{+}$ and $\mu_{-}$ ($0<\mu_{-}<\mu_{+}$) be the eigenvalues of $\vec{M}$. Then, the transition probability distribution can be expressed as \cite{Deng17b}
	\begin{equation}
		p(\Omega_{\tau}|\Omega_{0})=
		\begin{cases}
			\frac{1}{\pi\sqrt{(\Omega_{\tau}-\mu_{-}\Omega_{0})(\mu_{+}\Omega_{0}-\Omega_{\tau})}},~\Omega_{\tau}\in[\mu_{-}\Omega_{0}, \mu_{+}\Omega_{0}]\\
			0,~\text{otherwise}
		\end{cases}\,.
		\label{eq:transitionProbOscillator}
	\end{equation}
	Moreover, we have
	\begin{equation}
		\mu_{\pm}=Q^{*}\pm\sqrt{Q^{*2}-1}\,,
	\end{equation}
	and, therefore, we have $\mu_{+}\mu_{-}=1$ and $\mu_{+}+\mu_{-}=2Q^{*}$.
	
	We now have all the ingredients to compute the quantities in the generalized Jarzynski equality \eqref{eq:FT3} and the maximum work theorem \eqref{eq:Wmax}. A few simple lines of algebra reveal that
	\begin{equation}
		\la W\ra=\frac{1}{\beta}\left(Q^{*}\frac{\omega_{\tau}}{\omega_0}-1\right)\,.
		\label{eq:classicalwork}
	\end{equation}
	We observe that (as expected) the larger the average work is, the more nonadiabatic the process is\footnote{Note that Eq.~\eqref{eq:classicalwork} is identical to a high-temperature expansion of the quantum expression  $\la W\ra= \hbar (Q^{*}\omega_{\tau} - \omega_{0})/2\,  \coth \left(\beta\hbar \omega_0/2\right)$ \cite{Deffner16}. }.
	
	More elucidating is the resulting expression for the conditional partition function \eqref{eq:Zcond}. We obtain
	\begin{equation}
		\mc{Z}(B|A)=\frac{1}{\beta\hbar\omega_\tau}\,\frac{1}{Q^*}\,.
	\end{equation}
	Using the known expression for the canonical partition function $Z_B=1/\beta\hbar\omega_\tau$ we finally have
	\begin{equation}
		D\left[p_{B|A}\,||\,p^\mrm{eq}_B\right]=\lo{Q^*}\,.
	\end{equation}
	As alluded to above, the KL-divergence between the conditional and the canonical distribution measures the non-adiabaticity of the process. More specifically, for the parametric harmonic oscillator it is simply given by the logarithm of the measure of adiabaticity $Q^*$. It vanishes for adiabatic processes, $Q^*=1$, and is strictly positive otherwise, since $Q^*\geq 1$.
	
	\section{Conditional stochastic work in the exclusive framework}
	\label{sec:excWork}
	
	Before we conclude, we also briefly outline the treatment of the conditional stochastic work in the exclusive framework. In this paradigm the Hamiltonian is written \cite{Jarzynski07,Horowitz07} as
	\begin{equation}
		H(\Gamma; \alpha_t)=H^{(0)}(\Gamma)-\alpha_t\cdot\eta(\Gamma)\,,
	\end{equation}
	where $\eta(\Gamma)$ is the  extensive parameter conjugate to the external control parameter $\alpha$. Typically, it is assumed \cite{Horowitz07,Jarzynski07} that at $t=0$ we also have $\alpha_0=0$. 
	
	The standard exclusive work \cite{Bochkov1981,Horowitz07,Jarzynski07} is then defined as
	\begin{equation}
		W^{(0)}\equiv H^{(0)}(\Gamma_\tau)-H^{(0)}(\Gamma_0)\,.
	\end{equation}
	In complete analogy to above \eqref{eq:Wps} we define the conditional exclusive work as
	\begin{equation}
		\begin{split}
			W_\mrm{c}^{(0)}\left(\Omega_0^{(0)}\right)&\equiv \int d\Omega_\tau^{(0)} \,p\left(\Omega_\tau^{(0)}|\Omega_0^{(0)}\right) E^{(0)}\left(\Omega_\tau^{(0)}\right)-E^{(0)}\left(\Omega_0^{(0)}\right)\\
			&=\varepsilon^{(0)}\left(\Omega_0^{(0)}\right)-E^{(0)}\left(\Omega_0^{(0)}\right)\,,
		\end{split}
	\end{equation}
	where $\Omega^{(0)}$ is the phase space volume corresponding to the bare Hamiltonian $H^{(0)}(\Gamma)$. Correspondingly, the work distribution becomes
	\begin{equation}
		\mc{P}_\mrm{c}^{(0)}\left(W\right)=\int d\Omega^{(0)} \de{W-\varepsilon^{(0)}\left(\Omega_0^{(0)}\right)+E^{(0)}\left(\Omega_0^{(0)}\right)}\,p^\mrm{eq}\left(E^{(0)}\left(\Omega_0^{(0)}\right)\right)\,.
	\end{equation}
	
	It is then a simple exercise to show that we have
	\begin{equation}
		\la \e{-\beta W}\ra^{(0)}_\mrm{c}=\e{-D\left[p_{B|A}^{(0)}||p_B^{(0)}\right]}\,,
	\end{equation}
	where both, conditional thermal distribution as well as canonical distribution are taken with respect to the bare Hamiltonian.

	\section{Concluding remarks}
	\label{sec:conc}
	
	Stochastic thermodynamics has broadened the scope of thermodynamic notions far beyond their original inception. However, when building a novel framework conceptual ambiguity is often inevitable until a convincing consensus is achieved. More often than not, however, this consensus arises from a simple majority vote of the scientific community. For instance, the inclusive work has become the commonly accepted notion of thermodynamic work for single trajectories, despite the fact that the exclusive work was proposed almost two decades earlier. Thus, alternative notions for thermodynamic considerations might still need to be revealed  to provide additional insight into the conceptual underpinnings of the theory.
	
	In the present analysis we have proposed and discussed an alternative notion for thermodynamic work along single trajectories, which is based on the conditional expectation value of the change of energy. This notion of work is motivated by the one-time measurement approach from quantum thermodynamics, but is otherwise entirely classically justified. As main results we obtained a generalized Jarzynski equality and a tightened maximum work theorem that account for how non-adiabatic the process is.
	
	\appendix
	
	\section{Proof of Eq.~\eqref{eq:KL}}
	\label{sec:proofKL}
	We show the detailed derivation of Eq.~\eqref{eq:KL}. The Kullblack-Leiber (KL) divergence of the conditional thermal distribution with respect to the canonical state of the final configuration is
	\begin{equation}
		D\left[p_{B|A}\,||\,p^\mrm{eq}_B\right]= \int dE_B\int dE_A p_{B|A}(E_A,E_B)\ln\left(\frac{p_{B|A}(E_A,E_B)}{p^{\mrm{eq}}_B(E_B)}\right)\,.
	\end{equation}
	From Eq.~\eqref{eq:CondThermalDis}, we have
	\begin{equation}
		\begin{split}
			\int dE_A\int dE_B & p_{B|A}(E_A,E_B)\ln p_{B|A}(E_A,E_B)\\
			=&-\lo{\mathcal{Z}(B|A)}-\beta\int dE_A\int dE_B p_{B|A}(E_A,E_B)\varepsilon_B(E_A)\\
			&+\int dE_A\int dE_B p_{B|A}(E_A,E_B)\ln p(E_B|E_A)\,.
		\end{split}
	\end{equation}
	Here, we can explicitly write 
	\begin{equation}
		\begin{split}
			\int dE_A\int dE_B& p_{B|A}(E_A,E_B)\varepsilon_B(E_A)\\
			&=\int dE_A\int dE_B\int dE_B'\, p(E_B|E_A)\frac{\e{-\beta \varepsilon_B(E_A)}}{\mc{Z}(B|A)}E_B'p(E_B'|E_A)\\
			&=\int\!dE_A\!\int\!dE_B'\,p(E_B'|E_A) \frac{\e{-\beta \varepsilon_B(E_A)}}{\mc{Z}(B|A)}E_B'\\
			&=\int dE_A \int dE_B'\,p_{B|A}(E_A,E_B')E_B'\,,
		\end{split}
	\end{equation}
	where we used 
	\begin{equation}
		\int dE_B p(E_B|E_A)=1\,
	\end{equation}
	in the third line.
	Also, note that we have the vanishing conditional entropy
	\begin{equation}
		-\int dE_A\int dE_B p_{B|A}(E_A,E_B)\ln p(E_B|E_A) =0
	\end{equation}
	because the time evolution of the system is deterministic due to Eq.~\eqref{eq:liou} so that $E_B$ is a function of $E_A$ and vice versa. More precisely, by definition, we have $E_A=H(\Gamma_0;\alpha_0)$ and $E_B\equiv H(\Gamma_{\tau};\alpha_{\tau})$, where $\Gamma_t$ obeys the Liouville equation, so that $E_B$ is the function of $E_A$. 
	Therefore, we have 
	\begin{equation}
		\begin{split}
			\int dE_A\int dE_B & p_{B|A}(E_A,E_B)\ln p_{B|A}(E_A,E_B)\\
			=&-\lo{\mathcal{Z}(B|A)}  -\beta\int dE_A\int dE_B p_{B|A}(E_A,E_B)E_B\,.
		\end{split}
		\label{eq:1stterm}
	\end{equation}
	
	Furthermore, from $p^{\mrm{eq}}_B(E_B)=\e{-\beta E_B}/Z_B$, we have
	\begin{equation}
		\int dE_A\int dE_B p_{B|A}(E_A,E_B)\ln p^{\mrm{eq}}_{B}(E_B)=-\lo{Z_B}-\beta\int dE_A \int dE_Bp_{B|A}(E_A,E_B)E_B\,.
		\label{eq:2ndterm}
	\end{equation}
	Therefore, from Eqs \eqref{eq:1stterm} and \eqref{eq:2ndterm}, we have 
	\begin{equation}
		D\left[p_{B|A}\,||\,p^\mrm{eq}_B\right] = -\lo{\frac{\mathcal{Z}(B|A)}{Z_B}}\,,
	\end{equation}
	which proves Eq.~\eqref{eq:KL}.

	\section{Transition probability distribution in energy and phase space volume}
	\label{sec:TransProb}
	
	In this appendix we prove the equivalence of the conditional transition probability distribution in energy \cite{Jarzynski15} and phase space volume \cite{Deng17a} representation. We start be considering,
	\begin{equation}
		p(\Omega_\tau|\Omega_0)=\frac{\int d\Gamma_0\,\de{E(\Omega_0; \alpha_0)-H(\Gamma_0; \alpha_0)}\,\de{\Omega_\tau-\Omega(H(\Gamma_\tau(\Gamma_0); \alpha_\tau); \alpha_\tau)}}{\int d\Gamma_0\,\de{E(\Omega_0; \alpha_0)-H(\Gamma_0; \alpha_0)}}\,,
	\end{equation}
	and we recognize the denominator as density of states \cite{Jarzynski15} of the initial energy surface. In general, we have
	\begin{equation}
		\chi(E; \alpha) = \int d\Gamma\,\de{E(\Omega; \alpha)-H(\Gamma; \alpha)}=\frac{\pd \Omega}{\pd E}\,.
	\end{equation}
	Thus, we can also write \cite{Deng17a},
	\begin{equation}
		p(\Omega_\tau|\Omega_0)=\int d\Gamma_0\,\de{\Omega_\tau-\Omega(H(\Gamma_\tau(\Gamma_0); \alpha_\tau); \alpha_\tau)}\,\frac{\de{E(\Omega_0; \alpha_0)-H(\Gamma_0; \alpha_0)}}{\chi(E(\Omega_0; \alpha_0); \alpha_0)}\,.
	\end{equation}
	Now, we note that from the elementary properties of the Dirac-$\delta$ function we have
	\begin{equation}
		\de{\Omega_\tau-\Omega(H(\Gamma_\tau(\Gamma_0); \alpha_\tau); \alpha_\tau)}\chi(E_B; \alpha_\tau) =\de{E_B-H(\Gamma; \alpha_\tau)}\,,
	\end{equation}
	where we used that $E_B=E(\Omega_\tau; \alpha_\tau)$. Hence, we also have with $E_A=E(\Omega_0; \alpha_0)$ that
	\begin{equation}
		\label{eq:variab}
		p(\Omega_\tau|\Omega_0)\, \chi(E_B; \alpha_\tau)  = p(E_B|E_A)\,.
	\end{equation}
	Equation~\eqref{eq:variab} is then equivalent to 
	\begin{equation}
		\int dE_{B}\,p(E_B|E_A)f(E_B)=\int_{0}^{\infty}d\Omega_\tau\,p(\Omega_\tau|\Omega_0)f(E_B(\Omega_\tau; \alpha_\tau))\,.
	\end{equation}
	which holds for any arbitrary test function $f(E_B)$. In conclusion we have shown that the transition probability distributions in energy and phase space representation are, indeed, equivalent.

	\begin{acknowledgements}
		This work is supported by the U.S. Department of Energy, the Laboratory Directed Research and Development (LDRD) program and the Center for Nonlinear Studies at LANL. AS offers his gratitude to Yi-Xiang Liu and Paola Cappellaro for insightful discussions.
	\end{acknowledgements} 
	
	\section*{Conflict of interest}
	
	The authors declare that they have no conflict of interest.

	\bibliographystyle{spphys}    
	\bibliography{refOTM.bib}
	
\end{document}